\documentclass{ws-ijmpe}

\begin{document}

\markboth{N. Paar, T. Nik\v si\' c, D. Vretenar, and P. Ring}{Relativistic description of exotic collective
excitation phenomena in atomic nuclei}

%
\catchline{}{}{}{}{}
%

\title{RELATIVISTIC DESCRIPTION OF EXOTIC COLLECTIVE \\ 
EXCITATION PHENOMENA IN ATOMIC NUCLEI}

\author{\footnotesize N. Paar}

\address{Institut f\"ur Kernphysik, Technische Universit\"at
   Darmstadt, \\
   Schlossgartenstrasse 9, D-64289 Darmstadt, Germany\\
nils.paar@physik.tu-darmstadt.de}

\author{T. Nik\v si\' c, D. Vretenar}

\address{Physics Department, Faculty of Science, 
   University of Zagreb, Croatia}

\author{P. Ring}

\address{Physik-Department der Technischen Universit{\"a}t
   M{\"u}nchen, \\
   D-85748 Garching, Germany}

\maketitle

\begin{history}
\received{(received date)}
\revised{(revised date)}
\end{history}

\begin{abstract}
The low-lying dipole and quadrupole states in neutron rich nuclei, 
are studied within the fully self-consistent relativistic quasiparticle
random-phase approximation (RQRPA), formulated in the canonical
basis of the Relativistic Hartree-Bogoliubov model (RHB),
which is extended to include the density dependent interactions. In heavier
nuclei, the low-lying E1 excited state is identified as a pygmy dipole
resonance (PDR), i.e. as a collective mode of excess neutrons oscillating
against a proton-neutron core. 
Isotopic dependence of  the PDR is characterized by a crossing between
the PDR and one-neutron separation energies. 
Already at moderate proton-neutron asymmetry the PDR peak is
calculated above the neutron emission threshold, indicating important
implications for the observation of the PDR in ($\gamma$,$\gamma'$)
scattering, and on the theoretical predictions of the radiative neutron capture rates in 
neutron-rich nuclei. In addition, a novel
method is suggested for determining the neutron skin of nuclei, based
on measurement of excitation energies of the Gamow-Teller resonance
relative to the isobaric analog state.
\end{abstract}

\section{Introduction}
The multipole excitation phenomena in exotic nuclei, in particular the
properties of the low-energy excited states have raised significant
interest in many recent theoretical and experimental studies. 
 As one moves away
from the valley of $\beta$-stability
towards the neutron rich side, modification of the effective nuclear
potential leads to the formation of nuclei with diffuse neutron
densities, the occurrence of the neutron skin, halo
structures, and new modes of excitations. The weak binding of outermost
neutrons might give rise to the existence of  pygmy dipole resonance (PDR), 
when loosely bound neutrons coherently oscillate
against the proton-neutron core \cite{Suz.90,Cha.94}. The onset of
low-energy E1 strength has been observed in electromagnetic 
excitations in heavy-ion collisions in oxygen isotopes \cite{Lei.01},
and via $(\gamma,\gamma')$ scattering in lead
isotopes \cite{Rye.02,End.03} 
and N=82 \cite{Zil.02,Zil.04} isotone chain. Although the low-lying excitations
have been observed below or near the particle threshold, their structure
and collectivity
still remain under discussion \cite{Ada.96,Vrepyg.01,Paa.03,Sar.04}.
Recent study within the quasiparticle phonon model have shown 
that the low-lying excitations corresponding to the pygmy dipole resonance
are closely related to the size of the neutron skin \cite{Tso.04}. 
However, it has also been suggested that the low-lying E1 strength
could be of different origin \cite{Zil.02}. 
In addition to the two-phonon $2^+\otimes 3^-$ state, and the  
pygmy state, in this energy region one could also expect some
compressional low-lying isoscalar dipole strength,
maybe mixed with toroidal states \cite{Vre.02}, as well as the  
E1 strength generated by the breaking of the isospin symmetry 
due to a clustering mechanism \cite{Iac.84}. 
The low-lying excited states in weakly bound nuclei are well described by the
quasiparticle random phase approximation(QRPA) based on the HFB ground state.
In a recent study, the relativistic QRPA has been formulated
in the canonical basis of the relativistic Hartree-Bogoliubov model
 \cite{Paa.03}, and it has also been extended for the studies of 
charge exchange excitations \cite{PNVR.04}.
For understanding of the r-process, the
properties of Gamow-Teller ($J^{\pi}=1^{+}$) resonances(GTR) in
nuclei towards
the neutron drip-line are of a particular importance, since the half-lives
of $\beta$-decays along the r-process path are generally dominated by the 
low-energy tails
of GTR. The excitation energies of the spin-flip
and isospin-flip modes can also provide information on the neutron
skin of atomic nuclei \cite{Vre.03}.
\section{Self-consistent relativistic quasiparticle RPA in the relativistic
Hartree-Bogoliubov model}
In relativistic mean field models, the nucleus is described as a system
of Dirac nucleons that interact in a relativistic covariant manner by 
exchange of effective mesons. A quantitative theoretical description
of complex nuclear systems necessitates a density dependent interaction,
which is introduced via nonlinear $\sigma$ self-interaction, 
or by explicitely including phenomenological density dependence 
of $\sigma$, $\omega$, and $\rho$ meson-nucleon vertex functions, adjusted
to the properties of nuclear matter and finite nuclei \cite{Nik1.02}.  
The present investigation is based on the density dependent 
effective interaction DD-ME1 \cite{Nik1.02}.
In comparison to the interactions with non-linear $\sigma$-meson self-interaction
\cite{RRM.86,LKR.97},  the properties of asymmetric nuclear matter are much better described
by DD-ME1 interaction, and its improved isovector properties result
in better description of charge radii and neutron skin, what is essential
for the studies of exotic nuclei.
Description of the ground state in unstable open shell nuclei, 
characterized by the closeness of the Fermi surface to the particle continuum,
necessitates a unified description of mean-field and pairing correlations,
as for example in the framework of the Relativistic Hartree-Bogoliubov (RHB) theory.
The pairing correlations in RHB model are described by the finite range Gogny 
interaction D1S \cite{BGG.84}.
Excitations are studied within the relativistic quasiparticle random-phase
approximation (RQRPA) in the
configurational space, which is derived from the time-dependent RHB model
in the limit of small amplitude vibrations \cite{Paa.03,Nik2.02}.
The RQRPA equations are formulated in the canonical single-nucleon basis of 
RHB model which involves discretization of both the occupied states and 
the continuum.
In order to describe transitions to
low-lying excited states in weakly bound nuclei, the two-quasiparticle
configuration space must include states with both nucleons in the discrete
bound levels, states with one nucleon in the bound level and one nucleon in
the continuum, and also states with both nucleons in the continuum. 
In addition, a consistent treatment of the Dirac sea of negative energy states
is essential for successful application of RQRPA model.
The present RHB+RQRPA model is fully self-consistent, i.e. the same
effective interactions in $ph$ and
$pp$ channels, are used both in the RHB calculation for the ground state
and in RQRPA residual interaction, in order to recover the sum-rules and to
decouple the spurious states. 
\section{Low-lying multipole excitations in exotic nuclei}
As one moves towards nuclei away from the
valley of $\beta$ stability, in addition to giant resonances,
additional exotic excitations may appear in the low-energy region.
In the present study, the fully self-consistent RHB+RQRPA model,
based on interactions with density dependent couplings, is used to
investigate the properties of low-lying multipole excitations
as approaching towards exotic nuclei.
As  the neutron number increases along the isotope chain, the isovector dipole
strength distribution is characterized by its spreading into the
low-energy region, and by the mixing of isoscalar and isovector
modes \cite{Paa.03}.

In Table~1 RHB+RQRPA low-lying E1 strength $(E<15 MeV)$ is compared with
non-relativistic models: continuum linear response theory based on HFB ground state \cite{Mat.01}, shell
model calculations \cite{Sag.01}, and QRPA plus phonon coupling model
(with the neutron pairing gap $\Delta_n=12/{\sqrt A}~MeV$) 
 \cite{Col.01}, in comparison with the systematic experimental 
data from electromagnetic excitations in the heavy-ion collisions \cite{Lei.01}.
\begin{table}[pt]
\tbl{Integrated energy weighted dipole strength in oxygen isotopes up to
15 MeV in excitation energy, given in units of the TRK sum rule[\%].}
{\begin{tabular}{lcccc} \hline 
Oxygen isotope        A =        &  18  &  20  &  22  & 24 \\ \hline
RHB+RQRPA(DD-ME1)             &   11.7      &    11.9 &  15.2 & 20.9  \\ \hline 
continuum QRPA \cite{Mat.01}    &   7      &   11 &  16 & 21  \\ \hline 
Shell model \cite{Sag.01}     &   6.4      &    10.9 &  10.0 &  8.6  \\ \hline 
QRPA + phonon coupling \cite{Col.01}    &   7.0      &    9.0 & 
7.5 &    \\ \hline
Experiment (GSI) \cite{Lei.01}    &   8  &  12  &  7  &    \\ \hline
\end{tabular}}
\label{tableo}
\end{table}
For isotopes heavier than $^{20}$O  the agreement of RHB+RQRPA 
results with experiment is less satisfactory; partly it is due to the fact that in the RHB
model the drip-line for Z=8 is approached at $^{28}$O, instead $^{24}$O.
The low-energy dipole strength in light nuclei is composed from
the non-resonant single particle excitations of loosely bound
neutrons, and therefore it does not correspond to the collective
PDR mode \cite{Vrepyg.01}. 
However, the underlying structure of the low-lying dipole strength changes
with the mass number. In neutron-rich medium-heavy and heavy nuclei, 
a collective PDR state appears in the low-energy region due to vibration
of neutrons from the outer orbitals which give rise of the
neutron skin \cite{Vrepyg.01}.
In the present study, we extend investigation from Ref. \cite{Vrepyg.01}
by employing RHB+RQRPA model with density-dependent
interaction DD-ME1, which properly describes the size of neutron skin. 
\begin{figure}[th]
\vspace{0.4cm}
\centerline{\psfig{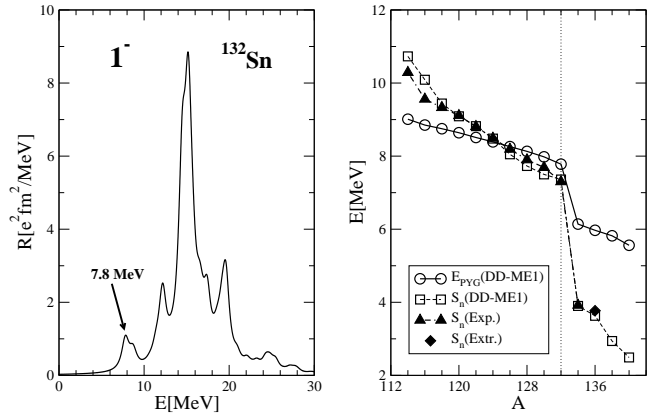}}
\caption{
RHB+RQRPA strength distribution of isovector dipole excitations 
in $^{132}Sn$, calculated with DD-ME1 effective interaction (left panel).
The PDR excitation energies and one-neutron separation energies are displayed as a
functions of the mass number for Sn isotopes, and compared with the
experimental and extrapolated values \protect\cite{Aud.95} (right panel).}
\label{figpyg1}
\end{figure}
In Fig.~\ref{figpyg1} we show the isovector dipole strength distribution in $^{132}Sn$.
In addition to the characteristic peak of the isovector giant
dipole resonance (IVGDR) at 15.2 MeV,
among several dipole states in the low-energy
region between 7 MeV and 10 MeV that are characterized by single particle
transitions, at  7.8 MeV a single pronounced peak is found
with a more distributed structure of
the RQRPA amplitude, exhausting 1.6$\%$ of
the energy weighted sum rule (EWSR), where EWSR=(1+$\kappa$)TRK. Here TRK
corresponds to the classical Thomas-Reiche-Kuhn sum rule \cite{Sag.01}, and the
enhancement factor from calculation equals $\kappa$=0.35.
The peak at 7.8 MeV is composed mainly of $11$ neutron $ph$ transitions from loosely
bound orbits,
each contributing more than $0.1 \%$ to the total RRPA amplitude  
$\sum_{\tilde{p} h}|X^{\upsilon}_{\tilde{p} h}|^2-
|Y^{\upsilon}_{\tilde{p} h}|^2=1, $
where $X^{\upsilon}$ and $Y^{\upsilon}$ are RRPA eigenvectors.
As we have shown in Refs. \cite{Vrepyg.01,Paa.03} by
analyzing the corresponding transition
densities, the dynamics of this low-energy mode is very different from that
of the IVGDR: the proton and neutron transition densities
are in phase in the nuclear interior,
there is almost no contribution from the protons
in the surface region, the isoscalar transition density dominates
over the isovector one in the interior, and
the large neutron component in the surface
region contributes to the formation of a node in the isoscalar
transition density. The low-lying pygmy
state does not belong to statistical E1
excitations sitting on the tail of the GDR, but represents a
fundamental structure effect: the neutron skin oscillates against the core.
In the right panel of Fig.~\ref{figpyg1}, RHB+RQRPA peak
energies of PDR 
are plotted as function of the mass number for Sn isotopes, in comparison
with the calculated one-neutron separation energies and
the corresponding experimental data and the extrapolated
value \cite{Aud.95}.  
One can observe that PDR excitation energy monotonously decreases 
with the mass number,
but a small kink appears at $N=82$ shell closure. The calculated
one-neutron separation energies of the Sn nuclei reproduce experimental
data in detail.
However, the separation energies decrease much faster than the 
calculated PDR excitation energies. At $N=82$, in 
particular, the separation energies display a sharp decrease, whereas
the shell closure produces only a weak effect on the PDR excitation 
energies. The important result here is that for $A < 124$ the PDR 
excitation energies are lower than the corresponding one-neutron 
separation energies, whereas for $A\geq 124$ the pygmy resonance
is located above the neutron emission threshold. This means, of course,
that in the latter case the observation of the PDR in $(\gamma,\gamma^\prime)$ 
experiments will be strongly hindered.
The relative position of the PDR with respect to the neutron emission
threshold will also have an effect on the calculated cross
sections for radiative neutron capture in neutron-rich Sn 
nuclei \cite{Gor.98,Gor.02}.

\begin{figure}[ht]
\vspace{0.55cm}
\centerline{\psfig{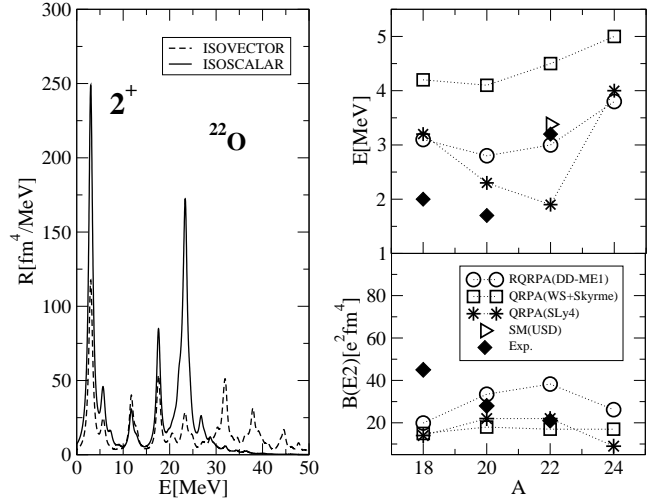}}
\caption{
RHB+RQRPA (DD-ME1) strength distribution of isovector and isoscalar quadrupole 
response in $^{22}O$ (left panel).
The calculated energies and B(E2) values of low-lying 2$^+$ states in oxygen isotopes 
are compared with nonrelativistic QRPA \protect\cite{Mat.02,Kha.02}, shell 
model calculatations \protect\cite{Thi.00} and experimental data \protect\cite{Ram.87,Bel.01}(right panel).}
\label{figquad}
\end{figure}

Another example how the exotic nuclear structure of loosely bound nucleons
may affect the properties of excitation phenomena, can be observed
in quadrupole excitations. The low-lying $2^+$ states in the
neutron rich oxygen isotopes have been already studied in the self-consistent
Skyrme Hartree-Fock-BCS + QRPA, demonstrating  that in exotic nuclei
the neutron to proton transition amplitudes for low-lying states differ 
noticeably from the simple N/Z estimate \cite{Kha.00}. The evolution of 
the low-lying $2^+$ strength strongly depends on the size of neutron
excess \cite{Mat.02}. 
The position of the
low-lying $2^+$ state is highly sensitive to the pairing correlations, 
and it is essential to include the pairing interaction in a fully
self-consistent way both in the ground state and in the residual RPA
interaction \cite{Mat.01,Paa.03}. 
In Fig.~\ref{figquad} we plot the isovector and isoscalar quadrupole
strength distribution in $^{22}$O.
The strong peak at $E \approx  23$ MeV in the isoscalar strength 
function corresponds to the isoscalar giant quadrupole resonance.
The low-lying state is located at 3.0 MeV, in
fair agreement with observed value at 3.2 MeV \cite{Bel.01}.
The isovector response, on the other hand, is strongly
fragmented, and distributed over the large region of excitation energies
$E\simeq 17-38$ MeV.  
In the right panel on Fig.~\ref{figquad} we compare the RHB+RQRPA energies
and transition probabilities B(E2) of the low-lying $2^+$ state in oxygen isotopes,  with results of previous studies.
The calculated  excitation energies of the first $2^+$ state  
in general slightly overestimate the measured excitation energies,
except in $^{22}$O. 
The RHB+RQRPA results for the low-lying $2^{+}$
excitation energies are qualitatively in agreement with
non-relativistic QRPA
calculations of the quadrupole response in neutron rich oxygen isotopes
 \cite{Kha.00,Kha.02,Mat.02}. The B(E2) values are obtained
above the other
theoretical predictions, and only in $^{20}$O the experimental
data is reproduced.
 
\section{Spin-isospin resonances and the neutron skin in nuclei}
The use of giant resonances is one among various experimental
methods that have been employed in the past to measure the size of the
neutron skin \cite{Kra.99}. 
In the following, a new method is suggested for determining the
difference between the radii of the neutron and proton density
distributions along an isotopic chain, based on
measurement of
the excitation energies of the Gamow-Teller resonances (GTR) relative
to the isobaric analog states (IAR) \cite{Vre.03}.
For the purpose of the present study, the RHB+RQRPA model has been extended
to treat the charge-exchange excitations - PN-RQRPA \cite{PNVR.04}. 
In addition to $\rho$-meson exchange in the residual interaction, the
pion with pseudo-vector type of meson-nucleon coupling is also included to
investigate unnatural parity excitations. 
Because of the derivative type of the pion-nucleon coupling, it is
also necessary to include a zero-range Landau-Migdal term that
accounts for the contact part of the nucleon-nucleon interaction.
It comes with a parameter $g^{\prime}= 0.55$, adjusted
to reproduce experimental data on the GTR excitation energies.
In the present analysis we also use the Gogny interaction
in the $T=1$ $pp$-channel of the PN-RQRPA.
For the $T=0$ proton-neutron pairing interaction in open shell nuclei
we employ a similar interaction: a short-range repulsive Gaussian
combined with a weaker longer-range attractive Gaussian \cite{PNVR.04}.
IAR strength distributions, i.e. the $J^{+}=0^{+}$ states, are
generated by the Fermi transition operator 
$T_{\beta^{\pm}}^{F}=\sum_{i=1}^{A}\tau_{\pm}$.
\begin{figure}[ht]
\vspace{0.3cm}
\centerline{\psfig{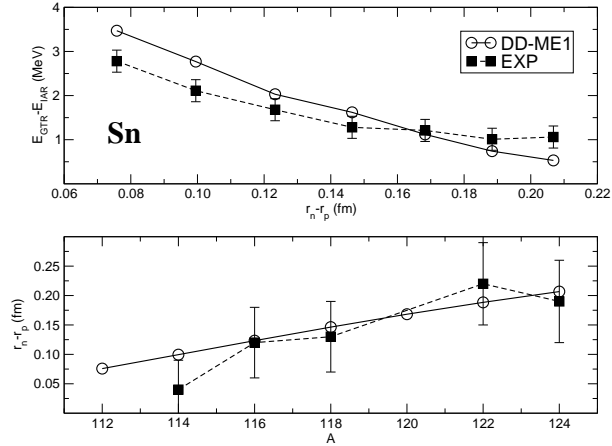}}
\caption{
The proton-neutron RQRPA and experimental \protect\cite{Pha.95}
differences between the excitation energies
of the GTR and IAR, as a function of the calculated 
differences between the rms radii of the neutron and proton 
density distributions of even-even Sn isotopes (upper panel). 
In the lower panel the calculated differences $r_n - r_p$
are compared with experimental data \protect\cite{Kra.99}.}
\label{figgt}
\end{figure}
On the other hand, GTR represents a coherent superposition of
high-lying $J^{\pi}=1^{+}$ proton-particle - neutron-hole configurations
of maximum collectivity associated with charge-exchange excitations
of neutrons from orbitals with $j = l + 1/2$ into proton
orbitals with $j = l - 1/2$. The GT operator reads, 
$T_{\beta^{\pm}}^{GT}=\sum_{i=1}^{A}\bm{\Sigma}\tau_{\pm} \; .$
In Fig.~\ref{figgt} we display the calculated differences between 
the centroids of the direct spin-flip GT strength and
the respective isobaric analog resonances for the
sequence of even-even Sn target nuclei. For $A=112 - 124$
the results of RHB plus proton-neutron RQRPA
calculation  (DD-ME1 density-dependent effective interaction,
Gogny $T=1$ pairing, $T=0$ pairing interaction \cite{PNVR.04} 
with $V_0 =$ 250 MeV, the Landau-Migdal parameter $g^{\prime}= 0.55$), 
are compared with experimental data \cite{Pha.95}.
As it has been emphasized in 
Ref. \cite{Vre.03}, the energy difference between the 
GTR and the IAS reflects the magnitude of the effective spin-orbit
potential. 
One can observe a uniform dependence of the energy 
spacings between the GTR and IAS on the size of the neutron-skin.
In principle, therefore, the value of $r_n - r_p$ can be determined
from the theoretical curve for a given value of 
$E_{\rm GT} - E_{\rm IAS}$. Of course, this necessitates  
implementation of a model which reproduces the experimental
values of the $r_n - r_p$, as it is displayed for RHB by using DD-ME1
effective interaction for Sn isotopes in the lower panel of Fig.~\ref{figgt}. 
\section{Conclusions}
In summary, the  
RHB plus relativistic QRPA which is extended to include the interactions
with density dependent meson-nucleon coupling constants (DD-ME1),
represents an advanced microscopic fully self-consistent 
model to investigate the properties of low-lying excitations in neutron-rich
nuclei. It is demonstrated that the one-neutron separation energies in Sn isotopes
decrease much faster with the mass number than PDR excitation energies, 
resulting with a characteristic crossing already at moderate neutron-proton
asymmetry. Finally, it is also shown that the energy  spacings between the GTR
and IAS provide direct information on the evolution of neutron skin-thickness
along the Sn isotopic chain.

\vspace{0.5cm}
\leftline{\bf Acknowledgments}

This work has been supported in part by the Bundesministerium
f\"ur Bildung und Forschung under project 06 MT 193, and by the
Gesellschaft f\" ur Schwerionenforschung (GSI) Darmstadt.
N.P. acknowledges support from the Deutsche
Forschungsgemeinschaft (DFG) under contract SFB 634.

\end{document}